\documentclass[prl,aps,twocolumn,superscriptaddress,longbibliography,floatfix]{revtex4-2}

\usepackage[
  colorlinks=true,
  urlcolor=blue,
  linkcolor=blue,
  citecolor=blue,
  bookmarks=false,
  pdftitle={Topological nodal line semi-metals},
  pdfauthor={Ankur Das}
]{hyperref}

\usepackage{import}
\usepackage{subfiles}
\usepackage{graphicx}
\usepackage{bm}
\usepackage{amsmath}
\usepackage{amssymb}
\usepackage{mathtools}
\usepackage{amsfonts}
\usepackage[all]{xy}
\usepackage{xspace}
\usepackage{accents}
\usepackage{stmaryrd}
\usepackage{tabularx}
\usepackage{extarrows}
\usepackage{float}
\usepackage{multirow}
\usepackage{makecell}
\usepackage[capitalise]{cleveref}
\usepackage[dvipsnames]{xcolor}
\usepackage[normalem]{ulem}
\usepackage{pdfpages}
\makeatletter
\AtBeginDocument{\let\LS@rot\@undefined}
\makeatother

\creflabelformat{equation}{#2\textup{#1}#3}

\begin{document}
\title{Topological nodal line semimetals with chiral symmetry}
	
\author{Faruk Abdulla}
\affiliation{Harish-Chandra Research Institute, A CI of Homi Bhabha National
Institute, Chhatnag Road, Jhunsi, Prayagraj (Allahabad) 211019, India}
\author{Ganpathy Murthy}
\affiliation{Department of Physics and Astronomy, University of Kentucky, Lexington, KY, USA}
\author{Ankur Das}
\email{ankur.das@weizmann.ac.il}
\affiliation{Department of Condensed Matter Physics, Weizmann
Institute of Science, Rehovot, 76100 Israel}

\begin{abstract}
Topological semimetals in three dimensions display band-touchings at points (Weyl or Dirac 
semimetals) or nodal lines in the Brillouin zone. Weyl semimetals can occur with
internal symmetries only (time-reversal ${\cal T}$, charge conjugation ${\cal C}$, and a product of the
two, called chiral/sublattice symmetry ${\cal S}={\cal T}{\cal C}$).  Nodal line semimetals possessing 
solely internal symmetries have only been classified abstractly, while those with SU(2) spin rotation or 
crystalline symmetries are known more explicitly. We show that chiral
symmetry classes that are topologically nontrivial in three dimensions (namely class AIII,
CII, CI, and DIII) always have a stable gapless phase, which is a topological nodal line
semimetal. Our classification differs from previous approaches and has direct implications for 
gapless surface states.
\end{abstract}	

\maketitle


Band topology
\cite{TKNN_1982,Haldane1983,KaneMele2005,BernevigHughesZhang2006,FuKane2007,
MooreBalents2007,Roy2009,FuKaneMele2007,HasanKane2010,Moore2010,Bansil_Lin_Das_2016} has emerged
as an important
organizing principle of condensed matter physics over the past several decades. Topological
insulators \cite{MooreBalents2007,FuKaneMele2007,Roy2009,Moore2010,franz2013,Bansil_Lin_Das_2016,
HasanKane2010, Ryu2010} are a striking consequence of band topology, possessing a bulk gap
but gapless surface/edge
states protected by the topology \cite{JackiwRebbi1976,Teo_2010}. More than a decade ago, it was
realized that gapless phases of matter, such as semimetals with robust band crossings, can
also be topological \cite{Murakami2007,Wan2011,Burkov_Balents_2011,Burkov2011}. The topology
associated with the gapless points is captured by enclosing them in a lower-dimensional surface
in $k$-space on which the Hamiltonian is completely gapped and can thus be described as a lower
dimensional topological insulator \cite{Ryu2010,ChiuTeoSchnyderRyu2016,Armitage2018}.
In three dimensions, there are four
kinds of topological semimetals: Weyl semimetals \cite{Murakami2007,Wan2011,Armitage2018},
Dirac semimetals \cite{Young2012,Wang2012, Wehling2014,Armitage2018}, nodal line (NL) semimetals
\cite{Burkov2011,Fang2016}, and the recently introduced nexus semimetals
\cite{Heikkila2015,Chang2017,das2020,das2022}. 

In what follows, we will always preserve translation symmetry but will distinguish between ``internal" symmetries - time reversal ${\cal T}, \text{charge conjugation } {\cal C}$, and their product, the chirality ${\cal S}={\cal T}{\cal C}$ - and other ``crystalline" symmetries which may be some combination of spin-rotation and/or spatial symmetries. One body of previous work has classified topological gapless states with internal symmetry by enclosing the singularity of the Green's function \cite{Horava_2005} $G(\omega,\bf{k})$ in a lower dimensional surface in the BZ, which has all the symmetries of the full Hamiltonian \cite{Zhao_Wang_2013,Matsuura_2013,Zhao_Wang_2014}. For example, if the symmetry class enjoys ${\cal T}$, then the enclosing surface must contain both ${\bf k}$ and $-{\bf k}$, that is, be centrosymmetric.  In some cases, the enclosing surface must be extended to one or two higher dimensions, with the internal symmetries  applying to the extra dimensions as well. The Hamiltonian restricted to the enclosing surface (with or without extra dimensions) should be fully gapped and in the same symmetry class as the original model. If this Hamiltonian has a nontrivial topology on the enclosing surface, the gapless state is protected by topology. 

Another body of work has focused on nodal line semimetals protected by crystalline symmetries
in addition to internal symmetries. Examples include topological nodal line semimetals (TNLSMs) 
protected by mirror reflection symmetry
\cite{Chiu2014, Chen2015TopologicalCM,Kim2015SurfaceSO, Yamakage2016, Bian2016, Bian2016n, Sun2017TopologicalNL,  Nie2019, Wang2021TopologicalNC}; TNLSMs protected by time-reversal, inversion, and SU(2)
spin rotation symmetries  \cite{Fang2015,Kim2015, Weng2015TopologicalNS, Yu2015TopologicalNS, Huang2016}; 
and TNLSMs in spin-orbit coupled systems protected by time-reversal, inversion, and 
nonsymmorphic  symmetries \cite{Fang2015,Schoop2016,Hong2018MeasurementOT,Meng2020ANH, Wang2021SpectroscopicEF}. 
Most recently, Kramers' nodal lines have been identified in  non-centrosymmetric crystals  possessing 
mirror or  roto-inversion symmetry
\cite{Xie2021KramersNL}.

As a consequence of nontrivial band touching in the bulk, TNLSMs have been predicted to
exhibit flat (drumhead) surface states \cite{Heikkil_2011, Kim2015, Weng2015TopologicalNS,
Yu2015TopologicalNS, Chen2015TopologicalCM, Kim2015SurfaceSO, Hong2018MeasurementOT}, in a region bounded
by the projection of the nodal loop on the surface Brillouin zone (BZ) that can be experimentally measured \cite{Hosen_2020,Bian2016n,wang2017,Lou_2018,Zhou2019,Belopolski_2019,Lv2021,Chen2021,Stuart2022,Gao2023}. 
However, surface states of TNLSMs protected by crystalline symmetries are not
necessarily robust because boundaries generically break 
crystalline symmetries \cite{Fang2015,Fang2016}. Thus, identifying such TNLSMs from their surface
states in experiments poses a challenge.

In this work, we identify a new way to classify 3D TNLSMs with 
internal symmetries, without invoking any crystalline symmetries except translations. We focus on symmetry classes  possessing  chiral symmetry, and show that the nodal loops
(NLs) can be characterized  by a 1D topological invariant on a loop that
``winds around" the nodal loop. The main difference from previous work \cite{Zhao_Wang_2013,Matsuura_2013,Zhao_Wang_2014}
is that the enclosing loop need not be centrosymmetric. The drumhead surface states in models that are nontrivial under our classification are not dependent on crystalline symmetry and thus are robust at open surfaces. 

\begin{figure}
     \centering
     \includegraphics[width=\columnwidth]{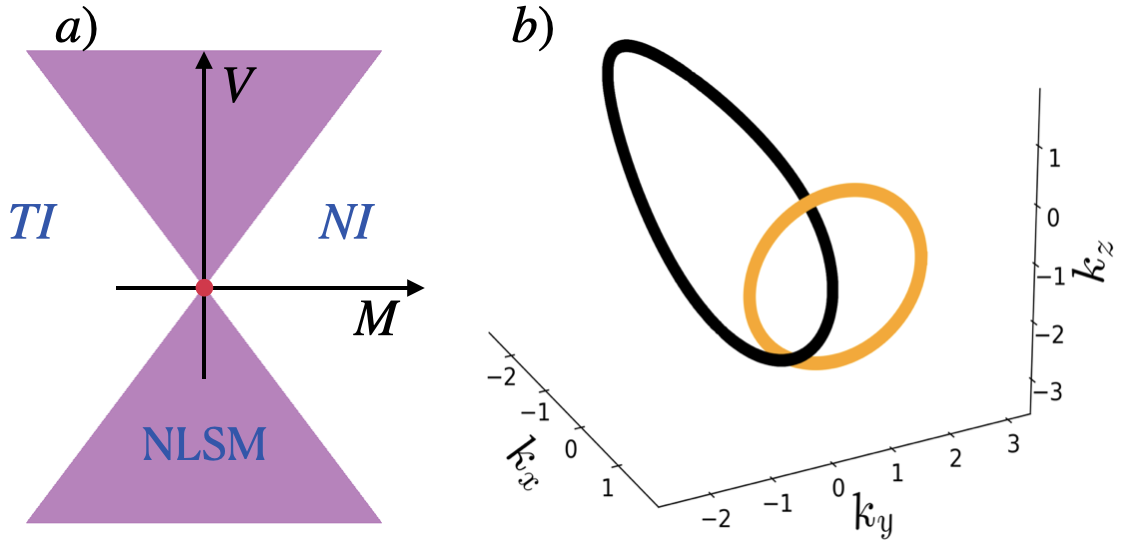}
     \caption{a) Schematically, the gapless phase that sits between the topological (TI) and normal 
     insulating (NI) states in the four chiral classes  AIII, CII, CI, and DIII is a topologically stable NLSM.
     b) A nodal loop (black) is enclosed by a 1D closed loop (orange) in 3D k-space. This nodal loop 
     is obtained by solving the AIII model in \cref{eq:genAIIIH}. The parameters are $A_1=A_2=A_3=V=\sqrt{2}$, 
     $M_0=0.4$, and $t^{(M)}_{ij}$ is taken diagonal $t^{(M)}_{11}=0$, $t^{(M)}_{22}=0.1$, and $t^{(M)}_{33}=0.2$.}               
 \label{fig:NLcartoon}
 \end{figure}

The starting point is to identify conditions under which a phase diagram must necessarily possess a gapless phase \cite{Murakami2007}. The four chiral Altland-Zirnbauer (AZ) \cite{Altland1997, Heinzner2005} classes  AIII, CII, CI and DIII all have nontrivial topology in 3D. The class BDI has chiral symmetry but is topologically trivial in 3D, and will be excluded henceforth. In the simplest models with the minimum number of symmetry-allowed couplings,
the transition from topological to trivial insulating states goes through a gapless Dirac point. Near this gapless point, if one adds all 1-body perturbations
consistent with the symmetry class, generically, an intermediate
{\it gapless phase} between the topological and trivial insulators emerges \cite{Murakami2007}, as shown schematically in \cref{fig:NLcartoon}a. In the chiral classes, we will show that this intermediate gapless phase is generically a TNLSM.

To see the relevance  of chiral symmetry to nodal lines, let us consider the following
two-band model in three dimensions belonging to the class AIII
\begin{align}\label{eq:chiralHamiltonian1}
    H({\bf k}) = f_1({\bf k})\eta^x + f_2({\bf k}) \eta^y,
\end{align}
where $\eta^x,\eta^y$ are Pauli matrices. 
We define time-reversal as ${\cal K}$ (complex conjugation of anything to its right), and the chirality operator as ${\cal S} = \eta^z$ satisfying  ${\cal S} H({\bf k}){\cal S}^{-1} = -H({\bf k})$. For a 
band touching we require $f_1({\bf k})=0$ and $f_2({\bf k})=0$. We have three variables
${\bf k} = (k_x, k_y, k_z)$, but only two conditions to satisfy. The solution space is
generically an NL (such as the black loop in \cref{fig:NLcartoon}b). To see
that the NL can be topologically protected in 3D, consider an arbitrary loop in the BZ (such as
the orange loop in \cref{fig:NLcartoon}b) that encloses the NL without
intersecting it. Along this loop, the 1D Hamiltonian is fully gapped (since the only gapless
points are assumed to occur on the NL), belongs to the symmetry
class AIII, and thus has a $Z$  classification \cite{Ryu2010,ChiuTeoSchnyderRyu2016}. The topological invariant
is the winding number given by
\begin{align}\label{eq:invariant}
    W = \frac{1}{2\pi i} \int_{0}^{2\pi} dk ~ \textrm{Tr}\left(Q^{-1} \partial_k Q\right).
\end{align}
The Hamiltonian in a chiral class can always be brought to block off-diagonal form, and
the matrix $Q(k)$ in \cref{eq:invariant} is the upper right block. If the winding number
of the 1D Hamiltonian is nonzero,  it is impossible
to shrink the loop to a point without hitting a singularity. In 3D ${\bf k}$-space,
this implies a line of singularities, that is, gapless points threading through the loop.
This example illustrates the possibility of a TNLSM with chiral symmetry (but no
crystalline symmetries) but is too simple to accommodate the full complexity of a generic 3D
Hamiltonian with spin and orbital degrees of freedom. 

Of the five chiral AZ classes,  AIII, CII, CI, and DIII have nontrivial topology in 3D. In each class, the simplest Hamiltonian describing the topological to trivial insulator transition can be written as a Dirac Hamiltonian, where changing the sign of the mass takes one through the transition.  As has been shown in previous work \cite{Murakami2007,Burkov2011}, adding all possible terms allowed by symmetry generically interposes a gapless phase between the topological and trivial insulators. This gapless 
phase may have topological protection, as was shown by Murakami \cite{Murakami2007} and 
Burkov and Balents \cite{Burkov2011} in class AII, where the gapless phase was found to be a topological Weyl 
semimetal. 

We now assert: (i) In classes  AIII, CII, CI, and DIII there is a generic gapless phase between the topological and trivial insulators. (ii) This gapless phase cannot generically be a Weyl semimetal. (iii) This gapless phase is generically a TNLSM where the nodal loops are classified by $Z$ invariant. We demonstrate assertion (i) by constructing the lowest (matrix) dimension model in each class (incorporating spin and orbital degrees of freedom) and computing the spectrum near the Dirac point  explicitly in the following. The physics behind assertion (ii) is the following: If an isolated Weyl node generically exists, it must be topologically protected. We surround 
it with an arbitrary, noncentrosymmetric, 2D enclosing surface. The 2D Hamiltonian 
restricted to this surface is only guaranteed to have chiral symmetry, since the surface is not centrosymmetric, and thus belongs to class AIII, which is trivial in 2D. Therefore the Weyl node is not protected. The gapless
phase cannot generically be a Dirac semimetal either, because the Dirac points are not protected without inversion 
or other crystalline symmetries.  Finally we come to assertion (iii). Suppose the gapless phase is a NL. We enclose it with an arbitrary 1D loop that winds around the nodal loop. The 1D
Hamiltonian restricted to this enclosing loop belongs to class AIII. Since class AIII has a $Z$ classification in 1D, the 1D Hamiltonian on the enclosing loop 
can be topologically nontrivial. If the winding number on the enclosing loop is nonzero, the nodal line has topological protection. 
Therefore our conclusion is that the gapless phase in classes  AIII, CII, CI and DIII is generically a TNLSM. In what follows, we will verify our  claim by writing down a generic Dirac Hamiltonian explicitly for each class, which we will then perturb with all symmetry-allowed perturbations and show that the gapless phase is indeed a TNLSM.

{\it Class AIII--} A generic spinful Hamiltonian in 3D belonging to class AIII
breaks time-reversal ${\cal T}$ and charge conjugation ${\cal C}$, but preserves their product
${\cal S}={\cal T} {\cal C}$. We start with a simple Hamiltonian in class DIII (${\cal T}^2=-1,\ {\cal C}^2=1$) showing the transition between the topological and trivial insulators.    
\begin{align}
    H_0({\bf k}) = \eta^x~{\bf k}. {\boldsymbol \sigma} + M~\eta^y.
\end{align}
Here $\sigma$'s and $\eta$'s are Pauli matrices acting on the spin and sublattice/chiral spaces.
We define 
and ${\cal T}=i \eta^x \sigma^y {\cal K}$, ${\cal C}=-\eta^y\sigma^y{\cal K}$, and  ${\cal S}=\eta^z$. We have assumed that all the low-energy degrees of freedom lie near the 
Gamma point in the BZ. This Hamiltonian has a number of ``accidental" symmetries in addition to ${\cal T}$ and ${\cal C}$, which we will
systematically break in order to ensure that the TNLSM we obtain does not have any
crystalline symmetry. For example, $H_0({\bf k})$ has inversion symmetry 
implemented by ${\cal I}=\eta^y$,
and mirror symmetries in the axes directions implemented by ${\cal M}_i=\eta^y \sigma^i$. 

There are seven $k$-independent terms that can be added to $H_0({\bf k})$ which break ${\cal T}$ and ${\cal C}$, but preserve
 ${\cal S}$. They consist of two vectors $ {\boldsymbol \alpha}= \{\eta^y\sigma^x, \eta^y\sigma^y, \eta^y\sigma^z\}$, $ {\boldsymbol \beta}= \{\eta^x\sigma^x, \eta^x\sigma^y, \eta^x\sigma^z\}$
and a scalar $\delta=\{\eta^x\}$. Adding a multiple of $\beta_i$ merely changes the position of the Dirac point, so we will ignore these terms. 
Adding an arbitrary linear combination of $\alpha_i$ and $\delta$, we obtain the most general Hamiltonian in class AIII,
\begin{align}\label{eq:genAIIIH}
 H_{\textrm{AIII}}({\bf k}) = H_0({\bf k}) + {\bf A}\cdot{\boldsymbol \alpha} + V \delta.
\end{align}
Note that $V\neq0$ is
essential in breaking inversion and mirror symmetries implemented by ${\bf 
A}\cdot{\boldsymbol \alpha}/|{\bf A}|$. 

Since the model in \cref{eq:genAIIIH} has to respect only the chiral symmetry to belong in class AIII, 
the parameters $M$, ${\bf A}$ and $V$ can be arbitrary functions of ${\bf k}$. We will assume that they 
are smooth near ${\bf k}=0$. The linear terms in ${\bf k}$ in these parameters will renormalize the matrix 
coefficients of ${\bf k}$ in $H_0({\bf k})$, and perhaps make the Fermi velocity angle dependent. We will ignore 
this effect. Thus, without loss of generality, we will assume the parameters to be quadratic
functions of ${\bf k}$: $M({\bf k})= M_0 - \sum_{ij} t^{(M)}_{ij} k_i k_j$, ~ $ V({\bf k})= V_0 - 
\sum_{ij} t^{(V)}_{ij} k_i k_j$, ${\bf A} = {\bf A}_0 - \sum_{ij} t^{({\bf A})}_{ij} k_i k_j$. In order 
to be consistent with our earlier assumption that all low-energy physics occurs near the Gamma point, 
we will need to restrict $t^{(\alpha)}_{ij}$, $\alpha=(M, V, {\bf A})$, to be a positive definite matrix in order to restrict the resulting nodal loop to the neighborhood of $k=0$ where our continuum Hamiltonian is valid. 
 By squaring $H_{\textrm{AIII}}$,
collecting terms, and squaring again, one obtains the spectrum.
\begin{equation}
    E^2={\bf k}^2+{\bf A}^2+V^2+M^2\pm2\sqrt{(V{\bf k}+M{\bf A})^2+({\bf k}\times{\bf A})^2}
\end{equation}
 The zero-energy condition is 
\begin{equation}
({\bf k}^2+M^2-{\bf A}^2-V^2)^2=-4({\bf k}\cdot{\bf A}-MV)^2.
\end{equation}
Clearly, the gapless points lie at the intersection of the two surfaces ${\bf k}^2+M^2 =
{\bf A}^2+V^2$ and ${\bf k}\cdot{\bf A}=MV$ in 3D k-space, and thus generically describe a nodal loop (NL). 
For ${\bf k}$-independent ${\bf A},~M,~V$, the spectrum has an accidental symmetry of rotation 
around the ${\bf A}$ direction. In this case, the NL lies at the intersection of the 
plane ${\bf k}\cdot{\bf A}=MV$ and the sphere ${\bf k}^2={\bf A}^2+V^2-M^2$, and is thus a 
planar circle.  However, for the generic $M= M_0-\sum_{ij}t^{(M)}_{ij}k_ik_j$ that we have 
assumed, there is no such accidental symmetry and the NL does not lie in a plane. 
An illustrative NL solution is presented in  
\cref{fig:NLcartoon}b for a representative choice of parameters. We have computed the winding 
invariant of \cref{eq:genAIIIH} (using \cref{eq:invariant}) around this NL numerically, 
which we find to be  $W=1$.  
This demonstrates our primary claim for class AIII: There is a  gapless phase which is generically 
a topologically protected NLSM. We reiterate that this phase does not need any crystalline/spin-rotation 
symmetries for its protection. Note 
that since a single nodal loop with a nonzero winding number can exist in class AIII, the only way 
to gap it out is to shrink it to a point.  

{\it Class CII--} A Hamiltonian in class CII has time-reversal, charge conjugation, and chiral 
symmetries,  with  ${\cal T}^2=-1,~{\cal C}^2=-1,~{\cal S}^2=1$, respectively. 
A minimal Dirac 
Hamiltonian including spin and orbital degrees of freedom is 8-dimensional \cite{Schnyder2008}. Near the topological to trivial transition, the
most general Dirac Hamiltonian close to the Gamma point (details in the companion long paper \cite{supp})
in class CII is 
\begin{equation}
\begin{aligned}
    H_{\textrm{CII}}({\bf k}) = & H_0({\bf k}) + \eta^y\otimes{\bf A}_a\cdot{\bf a}
    + A_5\eta^y\otimes\gamma^5
    \\&+ \eta^y \otimes {\bf V}_{b} \cdot {\bf b} + \eta^y \otimes 
    {\bf V}_{c} \cdot {\bf c}
     \\ &+ \eta^x \otimes {\bf V}_{d} \cdot {\bf d} +  V_{\epsilon} \eta^x \otimes \epsilon,
\end{aligned}
\label{eq:Ibreak}
\end{equation}
where, $H_0({\bf k}) = \eta^x \otimes D({\bf k})$, and $D({\bf k}) = { D^{\dagger}({\bf k})}
= \sum_{a=1}^{3} k_a  \gamma^a + M \gamma^4 + \mu \gamma^0$. $\eta^a$, $\tau^a$, and $\sigma^a$ are the Pauli matrices in the sublattice, orbital, and spin spaces respectively. The four-by-four $\gamma$ matrices  are represented in the spin and 
orbital basis i.e. $\gamma \equiv \tau \otimes \sigma $. We choose the following representation: 
$\gamma^1=\tau^z \sigma^x, ~  \gamma^2=\tau^z \sigma^y,  ~\gamma^3=\tau^y \sigma^0, 
~\gamma^4=\tau^x\sigma^0$. The remaining gamma matrices are defined as 
$\gamma^5=\gamma^1\gamma^2\gamma^3\gamma^4$ and $\gamma_{ij}=-\frac{i}{2}[\gamma^i, \gamma^j]$ for 
$i<j=1,2,..5$. Internal symmetries are realized  by ${\cal T}=-i\sigma^y {\cal K}$, ${\cal 
C}=i\eta^z\sigma^y {\cal K}$,  and  ${\cal S}={\cal TC}= \eta^z$ respectively.  The four by four  ${\bf a}$, ${\bf b}$, 
${\bf c}$, ${\bf d}$ and $\epsilon$ matrices 
are ${\bf a}=( \gamma^1,\gamma^2,\gamma^3 )$, ${\bf b}=(\gamma_{23}, \gamma_{31}, \gamma_{12})$,
${\bf c}= (\gamma_{15}, \gamma_{25}, \gamma_{35} )$, ${\bf d}=(\gamma_{14}, \gamma_{24}, 
\gamma_{34})$ and  $\epsilon=\gamma_{45}$ respectively. The $V$-coefficients multiply terms that break 
inversion (implemented by ${\cal I}=\eta^x \tau^x$). It is easily checked that the generic Hamiltonian 
breaks inversion and mirror symmetries. 

Time reversal forces the parameters $M$, ${\bf A}$, $A_5$, ${\bf V}_{b}$, ${\bf V}_{b'}$,
${\bf V}_{p}$, and ${\bf V}_{\epsilon}$  to be even functions of ${\bf k}$. Since there many possible perturbations, we will proceed in two steps. First, we set all the $V$ terms to zero, which restores inversion symmetry. In this special case we  
find that the zero energy solutions describe a gapless phase
which is generically a four-fold degenerate nodal loop semimetal. The NL carries a nontrivial 
winding number $W=2$ (computed using \cref{eq:invariant}). In the second step, we turn on one of $V$ terms (say $V_{\epsilon}$),
which breaks inversion and lifts the double degeneracy of the nodal loop to give a pair of two-fold degenerate nodal 
loops related by time-reversal symmetry, each carrying a nonzero winding number $W=1$. Since each
of the nodal loops carries a non-zero winding number, turning on the rest of the $V$ terms cannot gap out 
the NL immediately. Therefore we conclude that the gapless phase in class CII is generically a 
topologically protected NLSM. 

\begin{figure}
    \centering
    \includegraphics[width=\columnwidth]{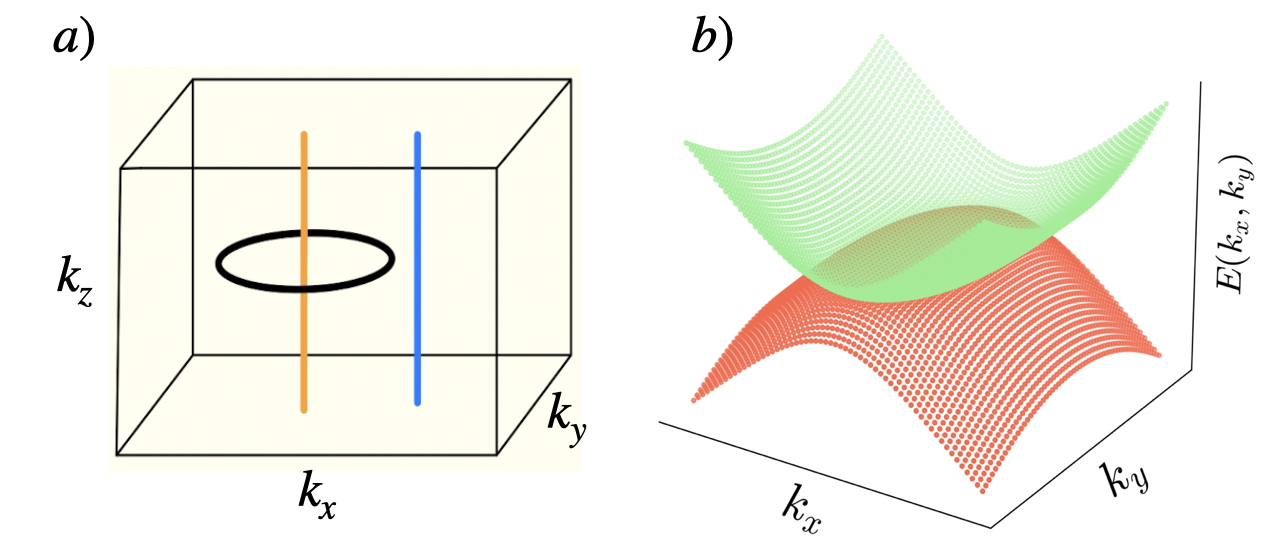}
    \caption{$a$) A nodal loop in the 3D BZ. 1D models in the AIII class, which is defined on a
     closed line will have non-zero winding invariant W when the closed line is inside the
     nodal loop. This leads to surface states for $k_x$-$k_y$ values bounded by the projection
     of the nodal loop on surface BZ as shown in cartoon figure $b$).}
    \label{fig:surface_states}
\end{figure}

{\it Class DIII and CI--} 
Hamiltonians in class DIII and CI have ${\cal T}^2=-1,~{\cal C}^2=1$ and  ${\cal T}^2=1,
~{\cal C}^2=-1$ respectively. Our calculations for class DIII/CI proceed very similarly to those of 
class CII (see the companion long paper Ref. \cite{supp} for details). As in classes CII and AIII, we find
a stable gapless phase, which is a TNLSM. In contrast to the CII case, where the pair of nodal loops
related by time-reversal symmetry carries the same winding number, in classes DIII/CI, they carry opposite 
winding numbers $W=\pm 1$. This is because when ${\cal T}^2=-{\cal C}^2$ the winding invariant $W$ of 
\cref{eq:invariant} changes sign under ${\cal T}$, whereas if ${\cal T}^2={\cal C}^2$, $W$ is invariant 
under ${\cal T}$. 
As a consequence, the transition between the topological/trivial insulator and the gapless TNLSM can occur
differently in classes AIII and CII on the one hand, and in classes CI and DIII on the other. 
In AIII/CII the only way to destroy nodal loops is to shrink them to zero, whereas in DIII/CI 
nodal loops related by ${\cal T}$ can annihilate in pairs and gap out. 

Thus, we have shown that in classes AIII, CII, CI, and DIII in $d=3$ there is a generic topologically 
protected nodal loop semimetal phase. Of course, TNLSMs may appear in contexts beyond this scenario, 
but in our approach, their existence is guaranteed. We provide a concrete algorithm for  constructing  
TNLSMs with no crystalline symmetries beyond translations. Further, we have provided a new classification 
for TNLSMs in models with chiral symmetry. The precise relation between our classification and the 
previous one \cite{Zhao_Wang_2013,Matsuura_2013,Zhao_Wang_2014} is not clear to us (see companion long 
paper \cite{supp} for more detail). The winding number we use is simpler to compute, and directly 
predicts the degeneracies of zero-energy drumhead surface states \cite{Heikkil_2011, Kim2015, 
Weng2015TopologicalNS, Yu2015TopologicalNS, Chen2015TopologicalCM, Kim2015SurfaceSO, Hong2018MeasurementOT,Hosen_2020,Bian2016n,wang2017,Lou_2018,Zhou2019,Belopolski_2019,Lv2021,Chen2021,Stuart2022,Gao2023}, 
as we now show.

A key experimental signature  \cite{Hosen_2020,Bian2016n,wang2017,Lou_2018,Zhou2019,Belopolski_2019,Lv2021,Chen2021,Stuart2022,Gao2023} of a TNLSM is the presence of gapless drumhead states in a region of the surface BZ \cite{Heikkil_2011, Kim2015, Weng2015TopologicalNS,
Yu2015TopologicalNS, Chen2015TopologicalCM, Kim2015SurfaceSO, Hong2018MeasurementOT}. 
Consider a nodal loop with a nonzero projected area on some plane in the BZ, such as the black
loop in \cref{fig:surface_states}a. Two test loops traverse the BZ perpendicular to
this surface, one inside the NL  projection (orange in \cref{fig:surface_states}a) and one
outside it (blue in \cref{fig:surface_states}a). The winding invariant
shows that these two 1D AIII Hamiltonians belong to different topological classes, implying the
presence of gapless states at the edge in the 1D model associated with the orange test loop, 
which implies gapless states in a region of the 2D BZ of the surface associated
with a given nodal loop. The degeneracies of drumhead states are slightly different in the four 
cases, and are presented in detail in the companion paper \cite{supp}.

Currently, it is not clear whether there are real materials with chiral symmetry, 
though there have been proposals of materials in classes DIII and AIII  \cite{Bauer_2004,Yuan_2006,Mondal2012}.
SrIrO${}_3$ \cite{Chen_2015}, and certain allotropes of Carbon \cite{Chen_2015_2} 
theoretically have chiral symmetry. However, these materials have nodal loops by virtue 
of crystalline symmetries, and so are not precisely what we are looking for.
Realizations of Hamiltonian in various chiral classes have been proposed in driven systems
\cite{Martin2017, Yang2018, Ozawa2019TopologicalQM} and in
optical lattices of ultracold atoms \cite{Song_2019,Schafer2020}. There has been some 
success in modeling fully gapped driven systems with nontrivial topology 
\cite{Long2021,Bhat2021OutOE,Esslinger_QHE_2023}. 

There are several open questions, the most pressing being the precise relation between our approach and the previous one \cite{Zhao_Wang_2013,Matsuura_2013,Zhao_Wang_2014} for class CII. Other important questions concern disorder and/or interactions, and the effects of orbital $B$ fields. The effect of disorder even on  the simpler 
point-node semimetals is currently not settled. While disorder-averaged field theoretic methods seem to perdict stability 
at small disorder strength \cite{Fradkin_1986_I,
Fradkin_1986, Goswami_etal_2011, Altland_2015,Roy_etal_2018,
Kobayashi_etal_2014,Sbierski_etal_2014,Louvet_Carpentier_Fedoreko_2016}, arguments based on the 
effects of rare Griffiths regions \cite{Rahul_2014, Pixley1_2016, Pixley2_2016, Pixley3_2017} suggest that they are immediately unstable.

For nodal loop semimetals, gapless modes are expected to be robust in the presence of disorder 
respecting chiral symmetry \cite{Matsuura_2013}. For generic disorder surface states are expected to decay algebraically 
into the bulk \cite{Silva_2023}, but persist till a critical bulk disorder strength \cite{PhysRevLett.124.136405}. The interplay of disorder and interactions has also been studied \cite{yejin_2016,Wang_Nandkishore_2017}.

The effect of orbital magnetic fields on NLSMs has been investigated for simple planar loops
\cite{Rhim2015,Alberto2018,Li2018,Yang2018_2,Molina2018,Chen2022}. As is known from
studies on WSMs \cite{abdulla_2022,Gooth_2023}, strong fields in the Hofstadter regime can produce
a rich phase structure. The behavior of generic loops in orbital fields remains an open question.  

We hope to address these interesting issues in the near future.

\begin{acknowledgements}
{\it Acknowledgments--}
The authors would like to thank Alexander Altland for very insightful discussions. The authors also thank Andreas P. Schnyder for his insights. FA would like to thank the Infosys Foundation for financial support and the ICTS for hospitality.
GM acknowledges the US-Israel Binational Science Foundation (grant no. 2016130), and the
hospitality of the Aspen Center for Physics (NSF grant no. PHY-1607611).  
A.D. was supported by the German-Israeli Foundation Grant No. I-1505-303.10/2019,
DFG MI 658/10-2, DFG RO 2247/11-1, DFG EG 96/13-1, CRC 183 (project C01),
and by the Minerva Foundation. A.D.
also thanks to the Israel Planning and Budgeting Committee (PBC) and the Weizmann Institute of
Science, the Dean of Faculty fellowship, and the Koshland Foundation for financial support.
The authors would like to thank the International Centre for Theoretical
Sciences (ICTS) for organizing the program -
Condensed Matter meets Quantum Information (code:
ICTS/COMQUI2023/9) and their hospitality where part of this work was completed.
\end{acknowledgements}

\bibliographystyle{apsrev}
\bibliography{export}

\end{document}